\documentclass[multphys]{svmult}
\usepackage{graphicx}

\begin{document}

\title{Polymer Simulations with a flat histogram stochastic growth algorithm}
\author{Thomas Prellberg\inst1\and
Jaroslaw Krawczyk\inst1\and
Andrew Rechnitzer\inst2}
\institute{
        Institut f\"ur Theoretische Physik,
        Technische Universit\"at Clausthal, 
        Arnold Sommerfeld Stra\ss e 6,
        D-38678 Clausthal-Zellerfeld, Germany,
\and
        Department of Mathematics and Statistics,
        The University of Melbourne,
        Parkville 3010, Australia.
}
\maketitle

\begin{abstract}
  We present Monte Carlo simulations of lattice models of polymers.
  These simulations are intended to demonstrate the strengths of a
  powerful new flat histogram algorithm which is obtained by adding
  microcanonical reweighting techniques to the pruned and enriched
  Rosenbluth method (PERM).
  

\end{abstract}

\section{Introduction}

Monte Carlo simulations of polymer models have played a significant role in
the development of Monte Carlo methods for more than fifty years
\cite{king1951}. We present here results of simulations performed with a
powerful new algorithm, flatPERM \cite{prellberg2004}, which combines a
stochastic growth algorithm, PERM \cite{grassberger1997}, with umbrella
sampling techniques \cite{torrie1977}. This leads to a flat histogram in a
chosen parameterization of configuration space.

The stochastic growth algorithm used is the pruned and enriched Rosenbluth
method (PERM) \cite{grassberger1997}, which is an enhancement of the
Rosenbluth and Rosenbluth algorithm \cite{rosenbluth1955}, an algorithm
that dates back to the early days of Monte Carlo simulations. While PERM
already is a powerful algorithm for simulating polymers, the addition of
flat-histogram techniques \cite{wang2001} provides a significant
enhancement, as has already been exploited in \cite{bachmann2003}, where it
has been combined with multicanonical sampling \cite{berg1991}.

Before we describe the algorithm in detail and present results of the
simulations, we give a brief motivating introduction to the lattice models
considered here.

If one wants to understand the critical behavior of long linear polymers in
solution, one is naturally led to a course-grained picture of polymers as
beads of monomers on a chain. There are two main physical ingredients
leading to this picture. First, one needs an ``\emph{excluded volume}''
effect, which takes into account the fact that different monomers cannot
simultaneously occupy the same region in space. Second, the quality of
the solvent can be modeled by an effective monomer-monomer interaction.
Monomers in a good solvent are surrounded by solvent molecules and hence
experience an effective monomer-monomer repulsion. Similarly, a bad solvent
leads to an effective monomer-monomer attraction.

\begin{figure}
\begin{center}
\includegraphics[width=7cm]{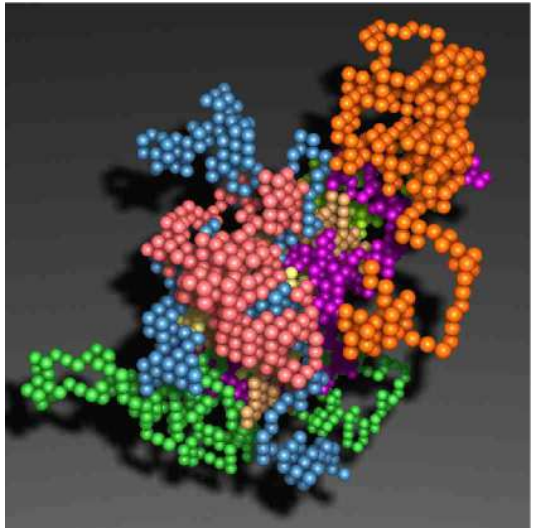}
\end{center}
\caption{
Eight lattice polymers in a bad solvent (picture courtesy of H. Frauenkron, FZ J\"ulich)
\label{figure1}
}
\end{figure}

Consequently, polymers in a good solvent form swollen ``\emph{coils}'',
whereas polymers in a bad solvent form collapsed ``\emph{globules}'' and
also clump together with each other (see Fig.~\ref{figure1}).  In order to
study the transition between these two states, it is advantageous to go to
the limit of an infinitely dilute solution, in which one considers
precisely one polymer in an infinitely extended solvent.

As we are interested in critical behavior, it is also possible to further
simplify the model by discretizing it. Due to universality, the critical
behavior is expected to be unchanged by doing so.  We therefore consider
random walks on a regular lattice, \emph{eg} the simple cubic lattice for a
three-dimensional model. One can think of each lattice site corresponding
to a monomer and the steps as monomer-monomer bonds.

We model excluded volume effects by considering {\em self-avoiding} random
walks which are not allowed to visit a lattice site more than once. The
quality of the solvent is modeled by an attractive short-range interaction
between non-consecutive monomers which occupy nearest-neighbor sites on the
lattice. At this point we may add more structure to our polymer model by
considering monomer-specific interactions. Specific properties of monomers
$i$ and $j$ on the chain lead to an interaction $\epsilon_{i,j}$ depending
on $i$ and $j$.

In this paper, we will consider three examples in detail. First, we
consider as pedagogical example, the problem of simulating polymers in a
two-dimensional strip. The interaction energy is simply $\epsilon_{i,j}=0$,
however, the introduction of boundaries makes simulations difficult
\cite{hsu2003b}.

\begin{figure}
\begin{center}
\includegraphics[width=7cm]{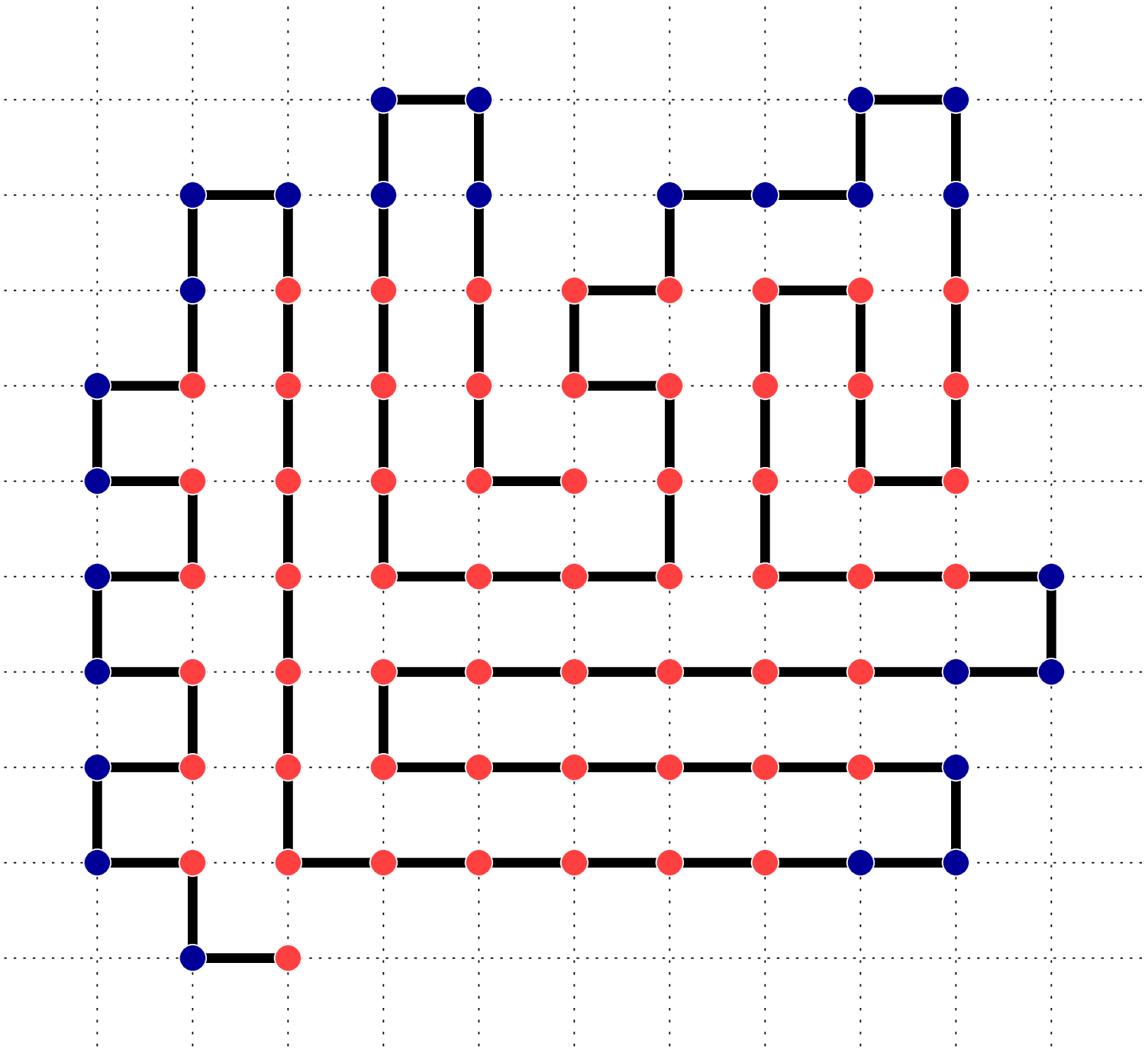}
\end{center}
\caption{HP model: shown is the conjectured groundstate of a sequence with
  85 monomers on the square lattice. The monomers with a lighter shade are
  of type H (hydrophobic), the monomers with a darker shade are of type P
  (polar).
\label{figure2}
}
\end{figure}

Our second example is the HP model which is a toy model of proteins
\cite{dill1985}.  It consists of a self-avoiding walk with two types of
monomers along the sites visited by the walk --- hydrophobic (type H) and
polar (type P). One considers monomer-specific interactions, mimicking the
interaction with a polar solvent such as water.  The interaction strengths
are chosen so that HH-contacts are favored, \emph{eg} $\epsilon_{HH}=-1$
and $\epsilon_{HP}=\epsilon_{PH}=\epsilon_{PP}=0$.  The central question is
to determine the density of states (and to find the ground state with
lowest energy) for a given sequence of monomers. An example of a
conjectured ground state is given in Fig.~\ref{figure2} for a particular
sequence of 85 monomers on the square lattice (the sequence is taken from
\cite{hsu2003}).

\begin{figure}
\begin{center}
\includegraphics[width=5cm]{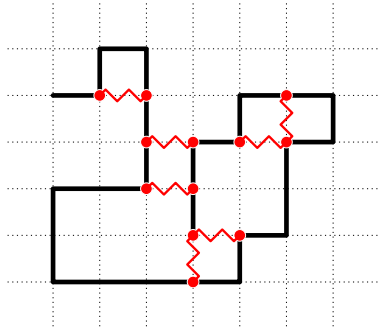}
\end{center}
\caption{A 26 step interacting self-avoiding walk on a square lattice with 7 interactions.
\label{figure3}
}
\end{figure}

Our third example is the interacting self-avoiding walk (ISAW) model of
(homo)-polymer collapse; it is obtained by setting $\epsilon_{i,j}=-1$
independent of the individual monomers. Here, one is interested in the
critical behavior in the thermodynamic limit, \emph{ie} the limit of large
chain lengths.  An example of an $26$-step interacting self-avoiding walk
with $7$ interactions is shown in Fig.~\ref{figure3}.

The partition function of $n$-step interacting self-avoiding walks can be written as
\begin{equation}
Z_n(\beta)=\sum_{\varphi}e^{-\beta E(\varphi)}=\sum_mC_{n,m}e^{\beta m}\;,
\end{equation}
where $E(\varphi)$ is the energy of an $n$-step walk, $\varphi$. Note that
the second sum is over the number $m$ of interactions, and $C_{n,m}$ is the
number of configurations of $n$-step self-avoiding walks with precisely $m$
interactions.

While the motivation for simulations of the various models is different,
the central problems turn out to be similar. For interacting self-avoiding
walks, the collapse transition is in principle understood. One has a 
tri-critical phase transition with upper critical dimension $d_u=3$,
so that one can derive the critical behavior from mean-field theory for
$d\geq3$ \cite{lawrie1984}, whereas for $d=2$ one obtains results 
from conformal invariance \cite{cardy1987}. However, even though this transition
is in principle understood, there are surprising observations above the upper
critical dimension \cite{prellberg2000}. Most importantly, there is no
good understanding of the collapsed regime, which is also notoriously
difficult to simulate.

Similarly, in the HP model one is interested in low-temperature problems,
\emph{ie} deep inside the collapsed phase. In particular, one wishes to
understand the design problem, which deals with the mapping of sequences
along the polymer chain to specific ground state structures. Again, the
most important open question is in the collapsed regime.

It is therefore imperative, to find algorithms which work well at low temperatures.
In the following section, we present just such an algorithm.

\section{The Algorithm}

This section describes our algorithm, as proposed in \cite{prellberg2004}.
The basis of the algorithm is the Rosenbluth and Rosenbluth algorithm, a
stochastic growth algorithm in which each configuration sampled is grown
from scratch. The growth is kinetic, which is to say that each growth step
is selected at random from all possible growth steps. Thus, if there are
$a$ possible ways to add a step then one selects one of them with
probability $p=1/a$.

\begin{figure}
\includegraphics[scale=0.45]{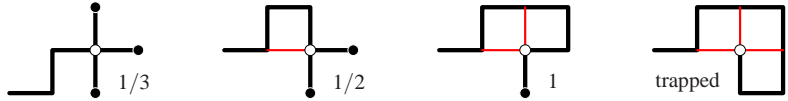}
\caption{
  For a self-avoiding walk on the square lattice, there can be between
  three and one ways of continuing, and the next step is chosen with equal
  probability from all possible continuations. In the right-most
  configuration, there is no way to continue, and the walk is trapped.
\label{figure4}
}
\end{figure}

For example, for a self-avoiding walk on the square lattice there may be
between one and three possible ways of continuing, but it is also possible
that there are no continuing steps, in which case we say that the walk is
\emph{trapped} (see Fig.~\ref{figure4}). 

As the number of possible continuations generally changes during the growth
process, different configurations are generated with different
probabilities and so one needs to reweight configurations to counter this.
If one views this algorithm as ``\emph{approximate counting}'' then the
required weights of the configurations serve as estimates of the total
number of configurations.

To understand this point of view, imagine that we were to perform a
\emph{complete} enumeration of the configuration space. Doing this requires
that at each growth step we would have to choose {\em all} the possible
continuations and count them each with equal weight. If we now select {\em
  fewer} configurations, then we have to change the weight of these
included configurations in order to correct for those that we have missed.
Thus, if we choose one growth step out of $a$ possible, then we must
replace $a$ configurations of equal weight by one ``representative''
configuration with $a$-fold weight. In this way, the weight of each grown
configuration is a direct estimate of the total number of configurations.

Let the \emph{atmosphere}, $a_n=a(\varphi_n)$, be the number of
distinct ways in which a configuration $\varphi_n$ of size $n$ can be
extended by one step. Then, the weight associated with a configuration
of size $n$ is the product of all the atmospheres $a_k$ encountered
during its growth, \emph{ie}
\begin{equation}
W=\prod_{k=0}^{n-1}a_k\;.
\end{equation}
After having started $S$ growth chains, an estimator $C_n^{est}$ for the
total number of configurations $C_n$
is given by the mean over all generated samples, $\varphi_n^{(i)}$, of size
$n$ with respective weights $W_n^{(i)}$, \emph{ie}
\begin{equation}\label{est}
C_n^{est}=\langle W\rangle_n=\frac1S\sum_iW_n^{(i)}\;.
\end{equation}
Here, the mean is taken with respect to the total number of growth 
chains $S$, and {\em not} the number of configurations which actually 
reach size $n$.  Configurations which get trapped before they 
reach size $n$ appear in this sum with weight zero.

The Rosenbluth and Rosenbluth algorithm suffers from two problems.
First, the weights can vary widely in magnitude, so that the mean may
become dominated by very few samples with very large weight.  Second,
regularly occurring trapping events, \emph{ie} generation of configurations with
zero atmosphere can lead to exponential ``\emph{attrition}'', \emph{ie}
exponentially strong suppression of configurations of large sizes.

To overcome these problems, enrichment and pruning steps have been
added to this algorithm, leading to the pruned and enriched Rosenbluth
method (PERM) \cite{grassberger1997}. The basic idea is that one wishes to
suppress large fluctuations in the weights $W_n^{(i)}$, as these should on
average be equal to $C_n$.

If the weight of a configuration is too large one ``\emph{enriches}'' by
making copies of the configuration and reducing the weights by an
appropriate factor. On the other hand, if the weight is too small, one
throws away or ``\emph{prunes}'' the configuration with a certain
probability and otherwise continues growing the configuration with a weight
increased by an appropriate factor. Note that neither $S$ nor the
expression (\ref{est}) for the estimate, $C_n^{est}$, are changed by either
enriching or pruning steps.

A simple but significant improvement of PERM was added in \cite{hsu2003a},
where it was observed that it would be advantageous to force each of the
copies of an enriched configuration to grow in \emph{distinct} ways. This
increases the diversity of the sample population and it is this version of
PERM that we consider below.

We still need to specify enrichment and pruning criteria as well as
the actual enrichment and pruning processes.  While the idea of PERM
itself is straightforward, there is now a lot of freedom in the
precise choice of the pruning and the enrichment steps. The ``art'' of
making PERM perform efficiently is based to a large extent on a
suitable choice of these steps --- this can be far from trivial!
Distilling our own experience with PERM, we present here what we
believe to be an efficient, and, most importantly, {\em parameter
  free} version.

In contrast to other expositions of PERM (\emph{eg} \cite{hsu2003}), we
propose to prune and enrich constantly; this enables greater exploration of
the configuration space. Define the \emph{threshold ratio}, $r$, as the
ratio of weight and estimated number of configurations,
$r=W_n^{(i)}/C_n^{est}$. Ideally we want $r$ to be close to $1$ to keep
weight fluctuations small. Hence if $r>1$ the weight is too large and so we
enrich. Similarly if $r<1$ then the weight is too small and so we prune.
Moreover, the details of the pruning and enrichment steps are chosen such
that the new weights are as close as possible to $C_n^{est}$:
\begin{itemize}
\item[$\bullet$]  $r>1$ $\rightarrow$ \textbf{enrichment step}:\\
  make $c=\min(\lfloor r\rfloor,a_n)$ distinct copies of the configuration,
  each with weight $\frac1cW_n^{(i)}$ (as in nPERM \cite{hsu2003a}).
\item[$\bullet$]  $r<1$ $\rightarrow$ \textbf{pruning step}:\\
  continue growing with probability $r$ and weight $\frac 1 r W_n^{(i)} =
  C_n^{est}$ (\emph{ie} prune with probability $1-r$).
\end{itemize}
Note that we perform pruning and enrichment {\em after} the configuration
has been included in the calculation of $C_n^{est}$. The new values are
used during the {\em next} growth step.

Initially, the estimates $C_n^{est}$ can of course be grossly wrong, as the
algorithm knows nothing about the system it is simulating. However, even if
initially ``wrong'' estimates are used for pruning and enrichment the
algorithm can be seen to converge to the true values in all applications we
have considered. In a sense, it is self-tuning.

We also note here, that the number of samples generated for each size is
roughly constant. Ideally, in order to effectively sample configuration
space, the algorithm traces an unbiased random walk in configuration size.
This means that PERM is, in some sense, already a flat histogram
algorithm. We shall return to this central observation below.

It is now straight-forward to change PERM to a thermal ensemble with
inverse temperature $\beta=1/k_BT$ and energy $E$ (defined by some
property of the configuration, such as the number of contacts) by
multiplying the weight with the appropriate Boltzmann factor $\exp(-\beta
E)$, which leads to an estimate of the partition function, $Z_n(\beta)$, of
the form
\begin{equation}
Z_n^{est}(\beta)=\langle W\exp(-\beta E)\rangle_n\;.
\end{equation}
The pruning and enrichment procedures are changed accordingly, replacing
$W$ by $W\exp(-\beta E)$ and $C_n^{est}$ by $Z_n^{est}(\beta)$. This gives
threshold ratio $r=W_n^{(i)}\exp(-\beta E^{(i)})/Z_n^{est}(\beta)$.  This
is the setting in which PERM is usually described.

Alternatively, however, it is also possible to consider microcanonical 
estimators for the total number $C_{n,m}$ of configurations of size $n$ 
with energy $m$ (\emph{ie} the ``density of states'').
An appropriate estimator $C_{n,m}^{est}$ is then given by the mean over
all generated samples $\varphi_{n,m}^{(i)}$ of size $n$ and energy $m$ 
with respective weights $W_{n,m}^{(i)}$, \emph{ie}
\begin{equation}
C_{n,m}^{est}=\langle W\rangle_{n,m}=\frac1S\sum_iW_{n,m}^{(i)}\;.
\end{equation}
Again, the mean is taken with respect to the total number of growth 
chains $S$, and {\em not} the number of configurations $S_{n,m}$ which 
actually reach a configuration of size $n$ and energy $m$. 
The pruning and enrichment procedures are also changed accordingly, 
replacing $C_n$ by $C_{n,m}$ and using
$r=W_{n,m}^{(i)}/C_{n,m}^{est}$.

As observed above, the pruning and enrichment criterion for PERM leads to
a flat histogram in length, \emph{ie} a roughly constant number of samples
being generated at each size $n$ for PERM. In fact, one can motivate the
given pruning and enrichment criteria by stipulating that one wishes to
have a roughly constant number of samples, as this leads to the algorithm
performing an unbiased random walk in the configuration size.  Similarly,
in the microcanonical version described above, the algorithm performs an
unbiased random walk in both size and energy of the configurations, and we
obtain a roughly constant number of samples at each size $n$ and energy
$m$.

It is because of the fact that the number of samples is roughly constant in
each histogram entry, that this algorithm can be seen as a ``flat
histogram'' algorithm, which we consequently call flat histogram PERM, or
flatPERM. In hindsight in becomes clear that PERM itself can be seen as a
flat histogram algorithm, at it creates a roughly flat histogram in size
$n$. Recognizing this led us to the formulation of this algorithm in the
first place.

We have seen that by casting PERM as an approximate counting method, the
generalization from PERM to flat histogram PERM is straight-forward and
(nearly) trivial. One can now add various refinements to this method if
needed. For examples we refer the reader to \cite{prellberg2004}. We close this
section with a summary of the central steps to convert simple PERM to
flatPERM by comparing the respective estimators and threshold ratios, $r$:

\begin{enumerate}
\item{\textbf{athermal PERM}:} estimate the number of configurations $C_n$
\begin{itemize}
\item[$\bullet$] $C_n^{est}=\langle W\rangle_n$
\item[$\bullet$] $r=W_n^{(i)}/C_n^{est}$
\end{itemize}
\item{\textbf{thermal PERM}:} estimate the partition function $Z_n(\beta)$
\begin{itemize}
\item[$\bullet$] $Z_n^{est}(\beta)=\langle W\exp(-\beta E)\rangle_n$
\item[$\bullet$] $r=W_n^{(i)}\exp(-\beta E_m^{(i)})/Z_n^{est}(\beta)$
\end{itemize}
\item{\textbf{flat histogram PERM}:} estimate the density of states $C_{n,m}$
\begin{itemize}
\item[$\bullet$] $C_{n,m}^{est}=\langle W\rangle_{n,m}$
\item[$\bullet$] $r=W_{n,m}^{(i)}/C_{n,m}^{est}$
\end{itemize}
\end{enumerate}
One can similarly generalize the above to several microcanonical
parameters, $m_1, m_2, \dots$, to produce estimates of
$C_{n,m_1,m_2,\dots}$.

Once the simulations have been performed the average of an observable, $Q$,
defined on the set of configurations can be obtained from weighted sums:
\begin{equation}
Q_{n,m}^{est}=\frac{\langle QW\rangle_{n,m}}{\langle W\rangle_{n,m}}=\frac{
\sum_iQ_{n,m}^{(i)}W_{n,m}^{(i)}}{\sum_iW_{n,m}^{(i)}}\;.
\end{equation}
These can then be used for subsequent evaluations. For instance, the
expectation value of $Q$ in the canonical ensemble at a given temperature
$\beta$ can now be computed via
\begin{equation}
Q_{n}^{est}(\beta)=\frac{\sum_m Q_{n,m}^{est}C_{n,m}^{est}\exp(-\beta E_m)}
{\sum_m C_{n,m}^{est}\exp(-\beta E_m)}\;.
\end{equation}
\emph{ie} only a single simulation is required to compute expectations
at \emph{any} temperature.

For many problems we are interested in their behavior at low temperatures
where averages of observables are dominated by configurations with high
energy. Such configurations are normally very difficult to obtain in
simulations. The flatPERM algorithm is able to effectively sample such
configurations because it obtains a roughly constant number of samples at
all sizes and energies (due to the constant pruning and enrichment). This
means that it is possible to study models even at very low
temperatures. Examples of this are given in the next section.

\section{Simulations}

A good way of showing how flatPERM works is to simulate two-dimensional
polymers in a strip. This kind of simulation has previously been performed
with PERM using Markovian anticipation techniques \cite{hsu2003b} which are
quite complicated. With flatPERM one simply chooses the vertical position
of the endpoint of the walk in the strip as an ``energy'' for the algorithm
to flatten against. We have found that this produces very good results.

\begin{figure}
\hspace*{-0.5cm}
\includegraphics[width=5.0cm,angle=-90]{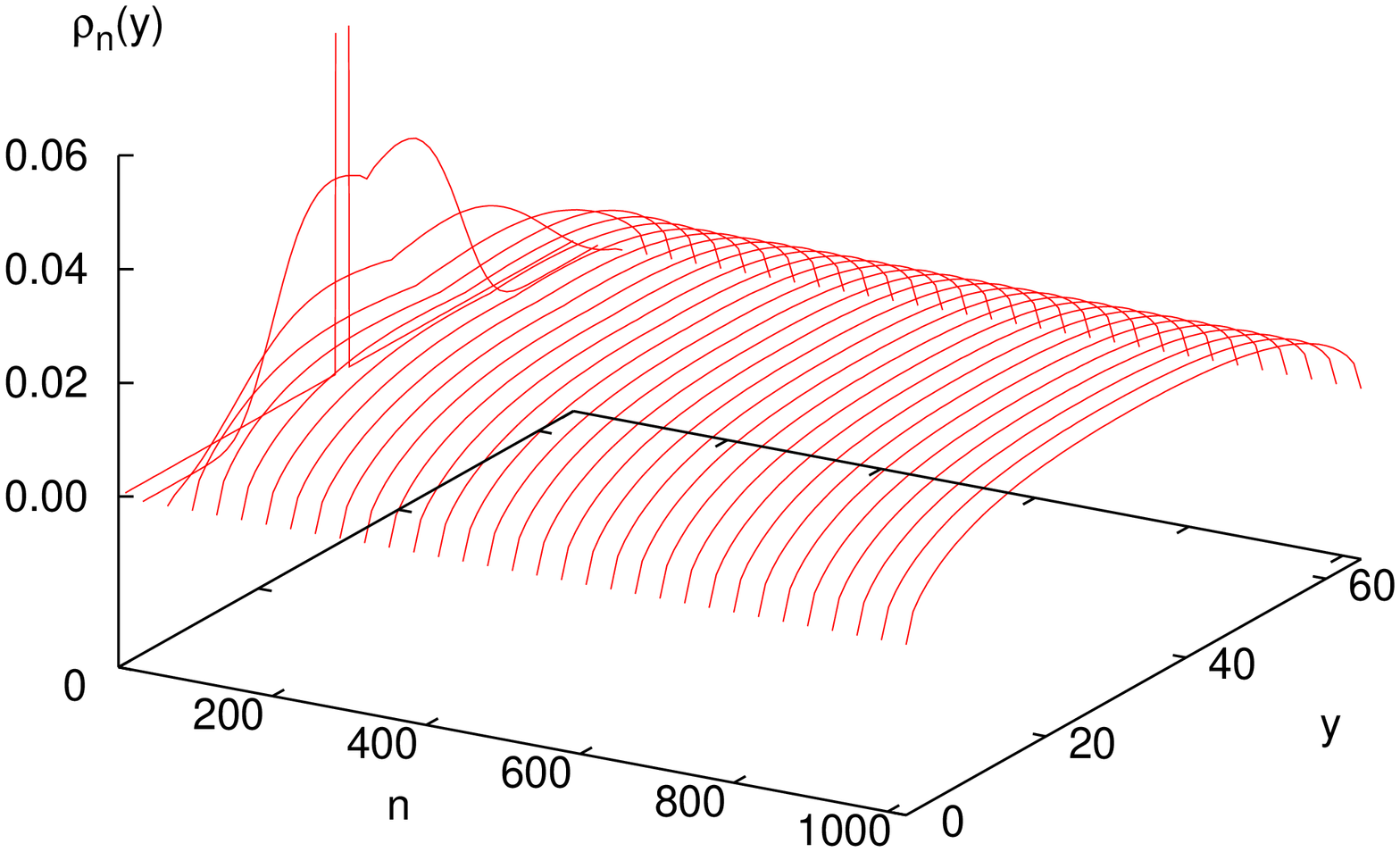}\hspace*{-1.0cm}
\includegraphics[width=5.0cm,angle=-90]{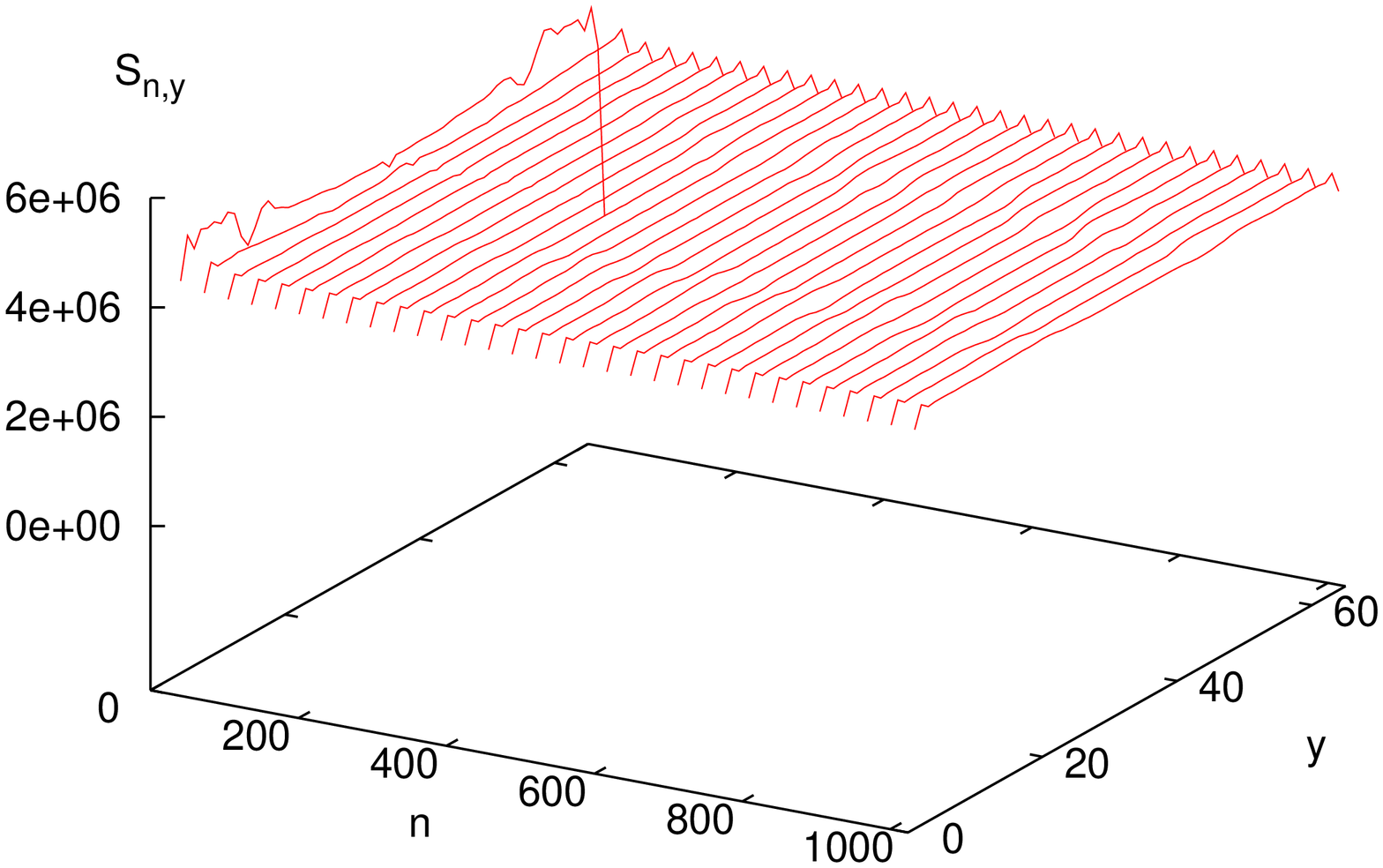}
\caption{Probability density $\rho_n(y)$ (left) and number of generated samples $S_{n,y}$ (right) versus
length $n$ and vertical endpoint coordinate $y$ for self-avoiding walk on a strip of width $64$ on
the square lattice.
\label{figure6}
}
\end{figure}

\begin{figure}
\begin{center}
\includegraphics[width=5.0cm,angle=-90]{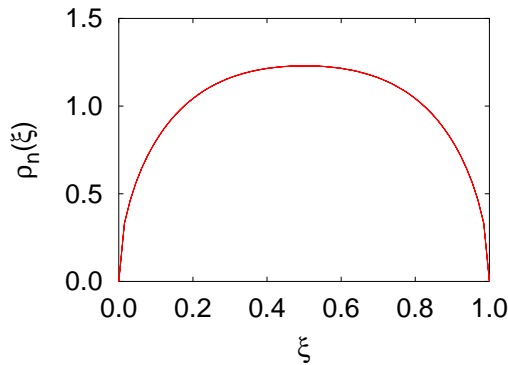}
\end{center}
\caption{Endpoint densities (scaled to the interval $[0,1]$) for lengths 512, 768, and 1024. The different
curves are indistinguishable.
\label{figure7}
}
\end{figure}

Fig.~\ref{figure6} shows the results of our simulations of $1024$-step
self-avoiding walks in a strip of width $64$. The left-hand figure is
the probability density $\rho_n(y)$ of the endpoint coordinate $y$
shown as a function of walk length $n$. The right-hand figure shows
the actual number of samples generated for each length $n$ and end
point position $y$.  One sees that the histogram of samples is indeed
nearly completely flat. One can now extract several quantities from
such simulations (see \cite{hsu2003b}), but we restrict ourselves here
to the scaled end-point density shown in Fig.~\ref{figure7}.

\begin{figure}
\begin{center}
\hspace*{-0.2cm}\includegraphics[width=8.3cm]{N14.1.sim.epsi}\\
\includegraphics[width=8.0cm]{N14.1cv.epsi}
\end{center}
\caption{
Sequence I (14 Monomers, HPHPHHPHPHHPPH, $d=3$): density of states versus energy (above) and specific heat $C_V/n$ versus
temperature $T$ (below).
\label{figure8}
}
\end{figure}

\begin{figure}
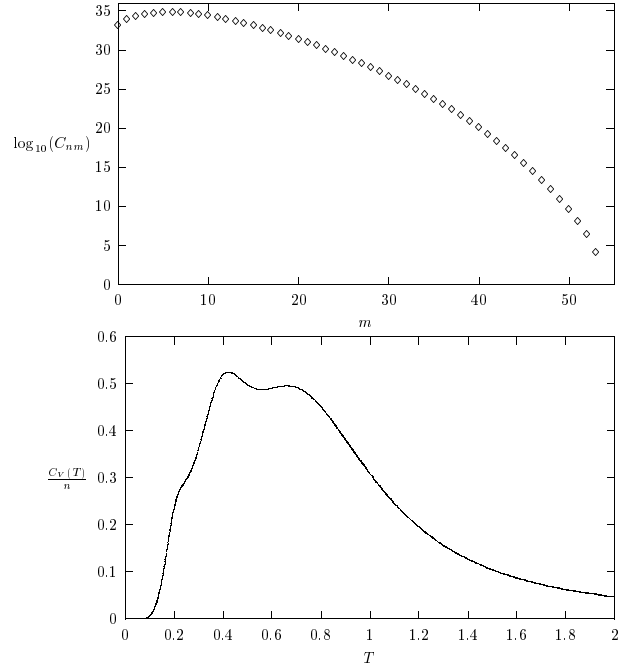

\begin{center}
\hspace*{-0.5cm}\includegraphics[width=8.0cm]{N85Cnm_1.epsi}\\
\includegraphics[width=7.6cm]{N85cv_1.epsi}
\end{center}
\caption{
Sequence II (85 Monomers, $d=2$): density of states versus energy (above) and specific heat $C_V/n$ versus
temperature $T$ (below).
\label{figure9}
}
\end{figure}

Next we show results from simulations of the HP-model. Here, we have
obtained the whole density of states for small model proteins with
fixed sequences. The first sequence considered (taken from
\cite{bachmann2003}) is small enough to enable comparison with exact
enumeration data. It has moreover been designed to possess a unique
ground state (up to lattice symmetries).

Fig.~\ref{figure8} shows our results. We find (near) perfect agreement
with exact enumeration even though the density of states varies over a
range of eight orders of magnitude! The derived specific heat data
clearly shows a collapse transition around $T=0.45$ and a sharper
transition into the ground state around $T=0.15$.

The next sequence (taken from \cite{hsu2003}) is the one for which
Fig.~\ref{figure2} shows a state with the lowest found energy.
Fig.~\ref{figure9} shows our results for the density of states and specific
heat. We find the same lowest energy $E=53$ as \cite{hsu2003} (though this
is not proof of it being the ground state).  The density of states varies
now over a range of 30 orders of magnitude! The derived specific heat data
clearly shows a much more complicated structure than the previous example.

For several other sequences taken from the literature we have confirmed
previous density of states calculations and obtained identical ground state
energies. The sequences we considered had $n=58$ steps ($3$ dimensions,
$E_{min}=-44$) and $n=85$ steps ($2$ dimensions, $E_{min}=-53$) from
\cite{hsu2003}, and $n=80$ steps ($3$ dimensions, $E_{min}=-98$) from
\cite{frauenkron1998}.  We studied also a particularly difficult sequence
with $n=88$ steps ($3$ dimensions, $E_{min}=-72$) from \cite{beutler1996},
but the lowest energy we obtained was $E=-69$.  While we have not been able
to obtain the ground state, neither has any other PERM implementation (see
\cite{hsu2003}).

\begin{figure}
\hspace*{-0.5cm}\includegraphics[width=5.0cm,angle=-90]{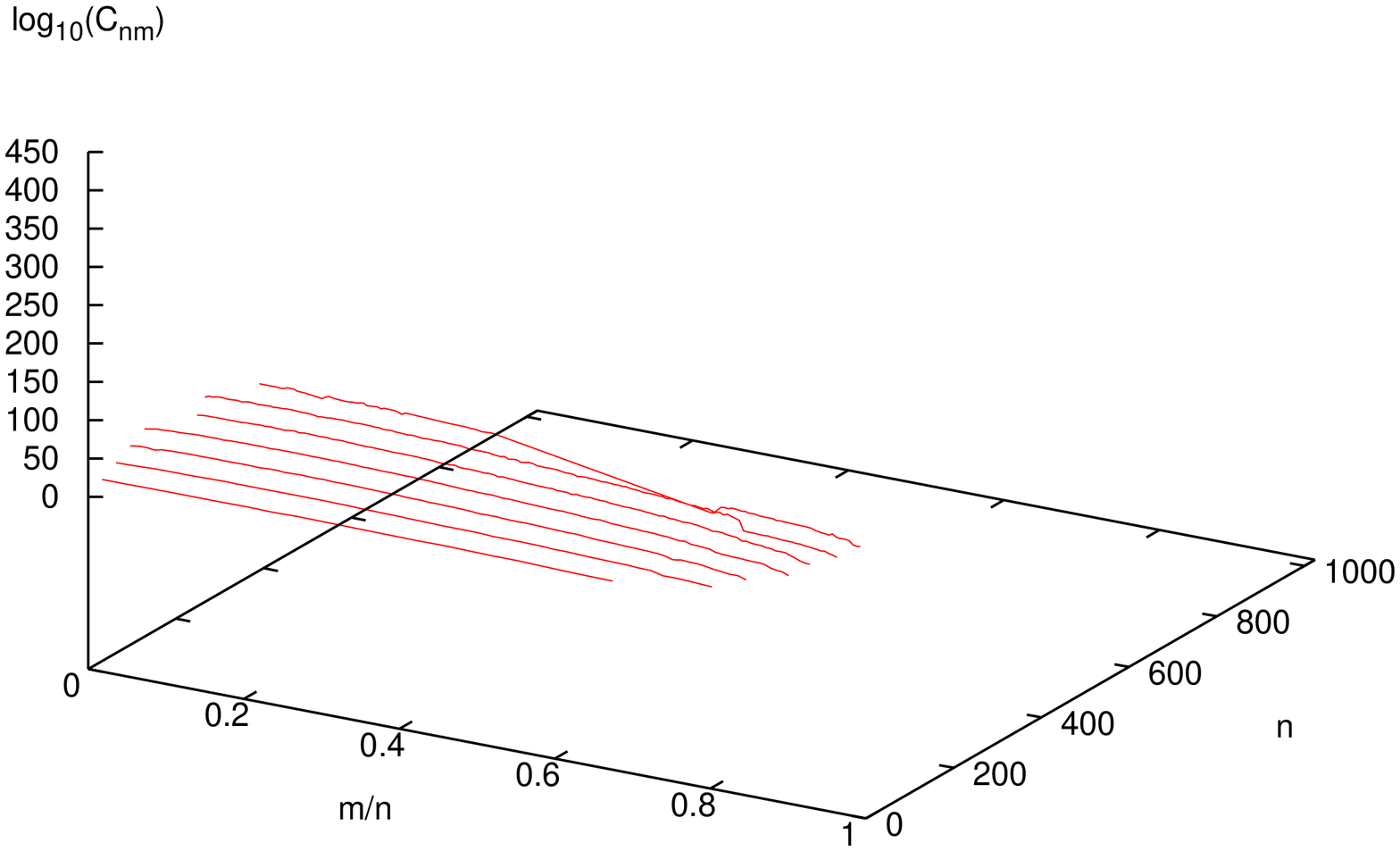}\hspace*{-1.5cm}
\includegraphics[width=5.0cm,angle=-90]{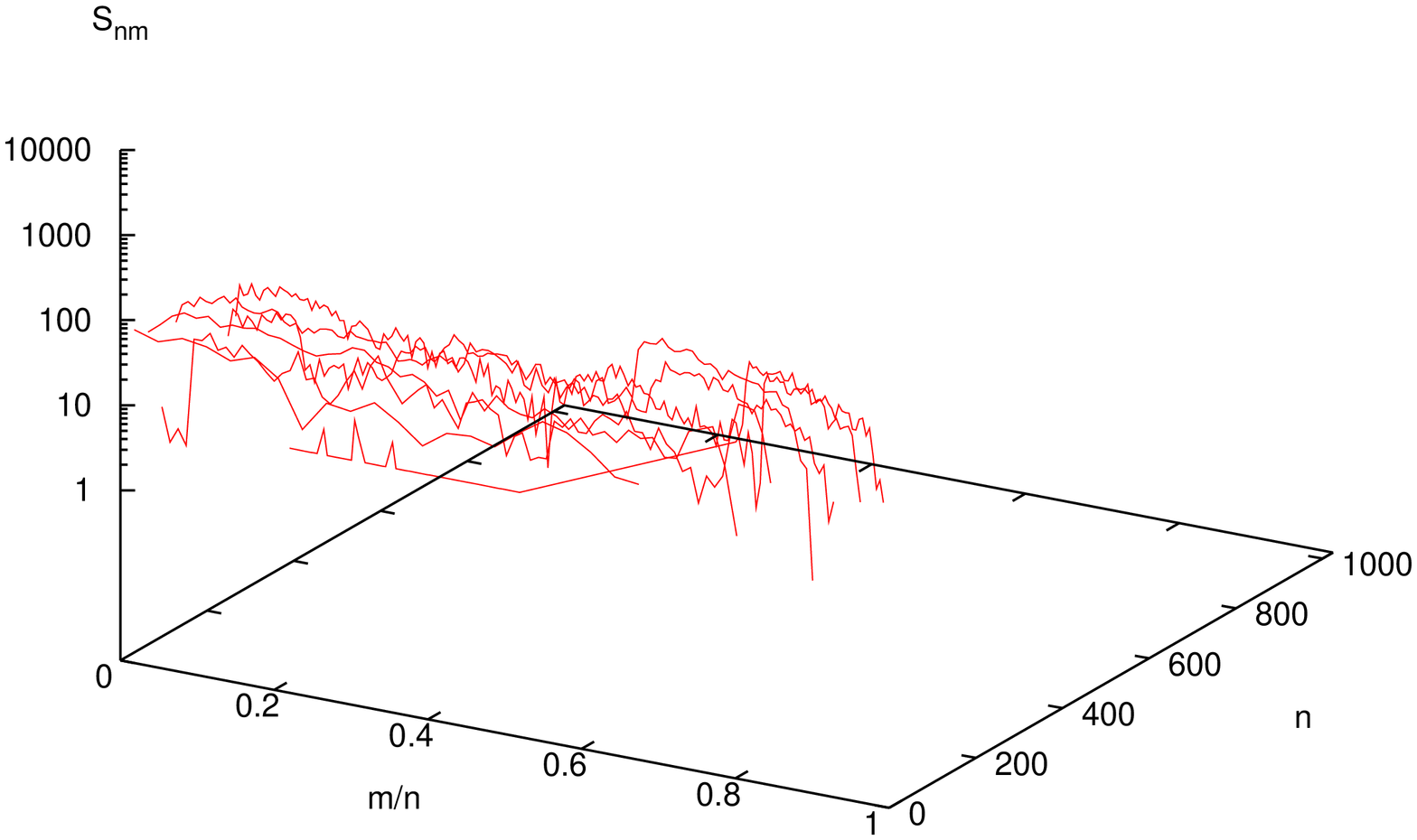}\vspace*{-1cm}\\
\hspace*{-0.5cm}\includegraphics[width=5.0cm,angle=-90]{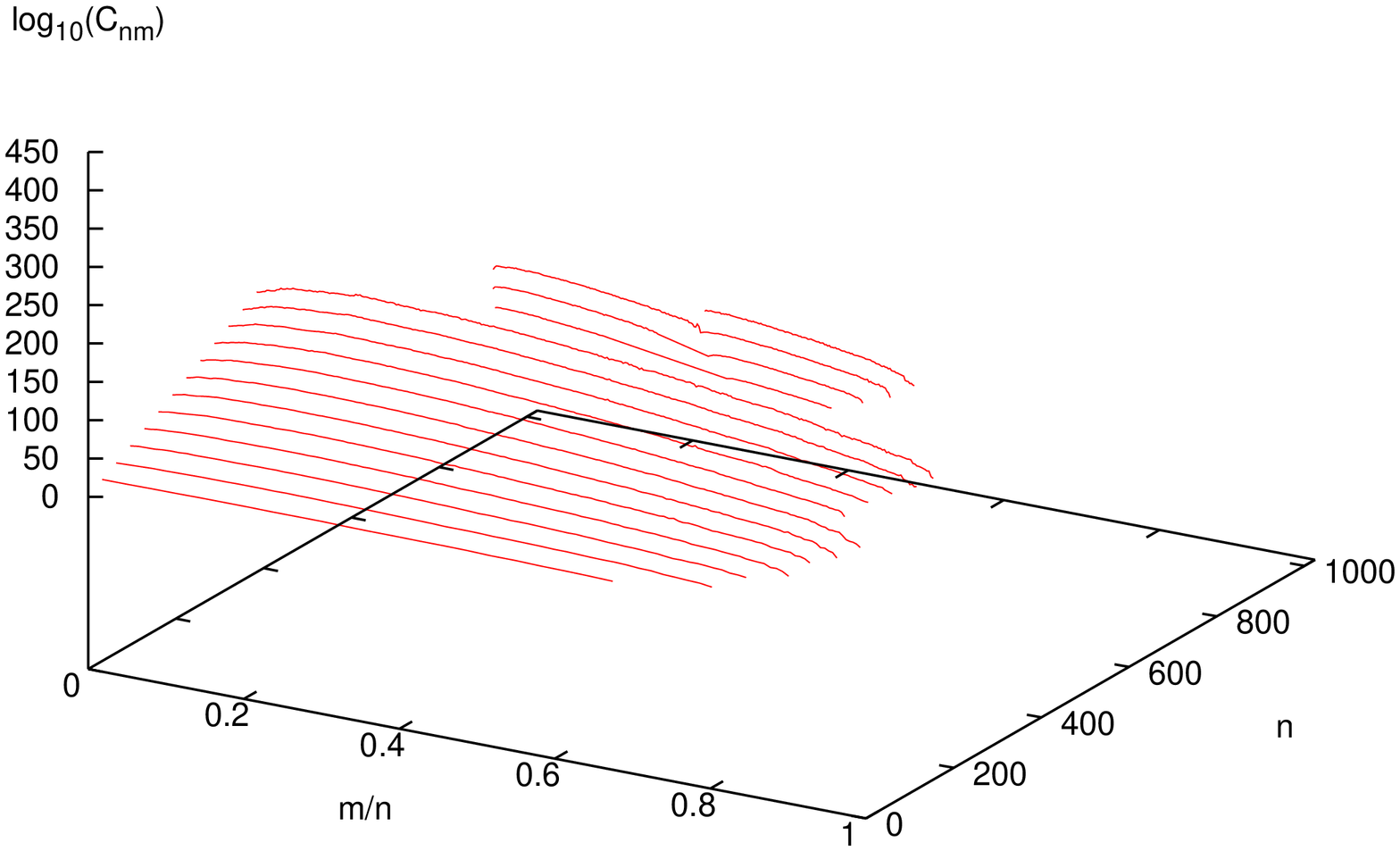}\hspace*{-1.5cm}
\includegraphics[width=5.0cm,angle=-90]{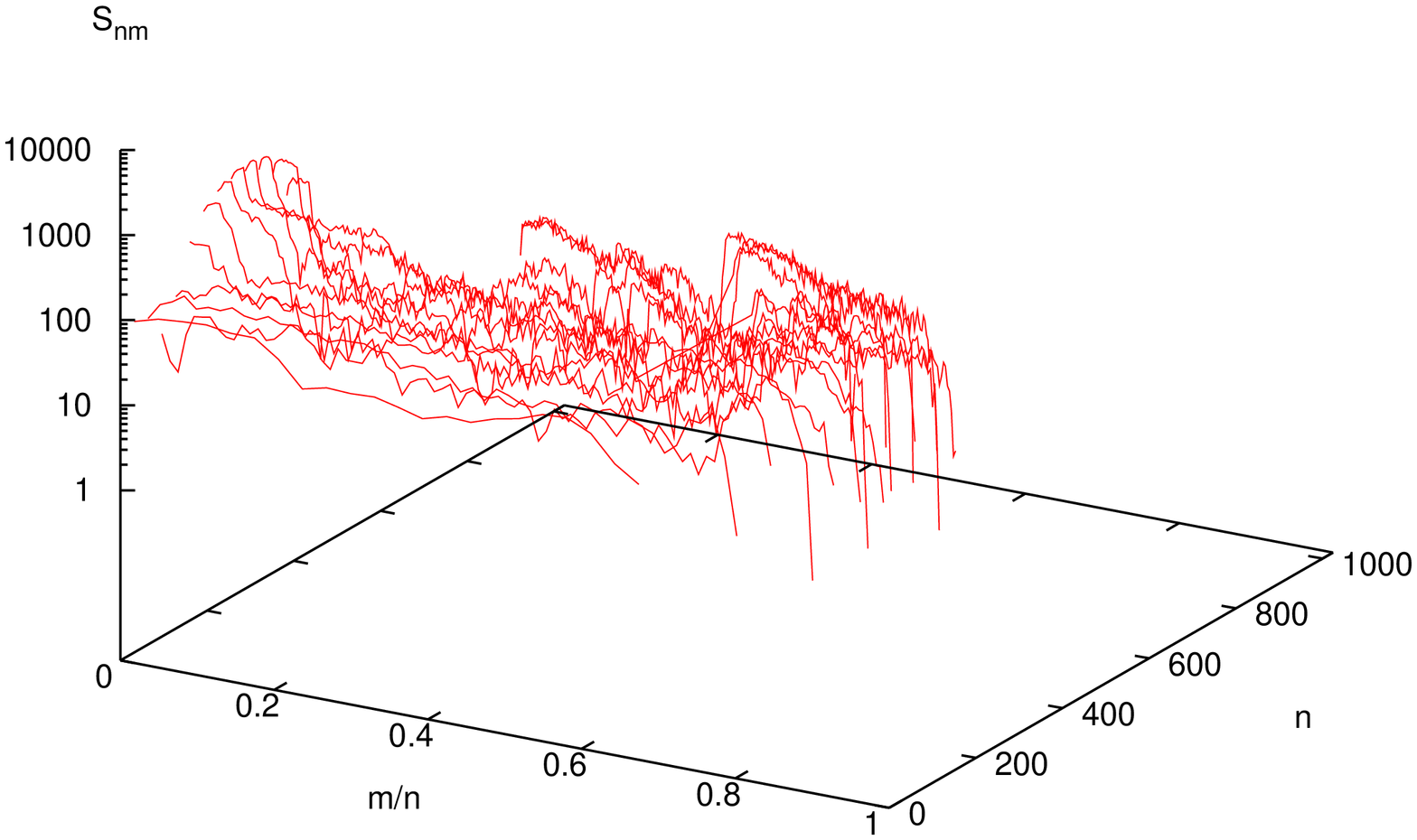}\vspace*{-1cm}\\
\hspace*{-0.5cm}\includegraphics[width=5.0cm,angle=-90]{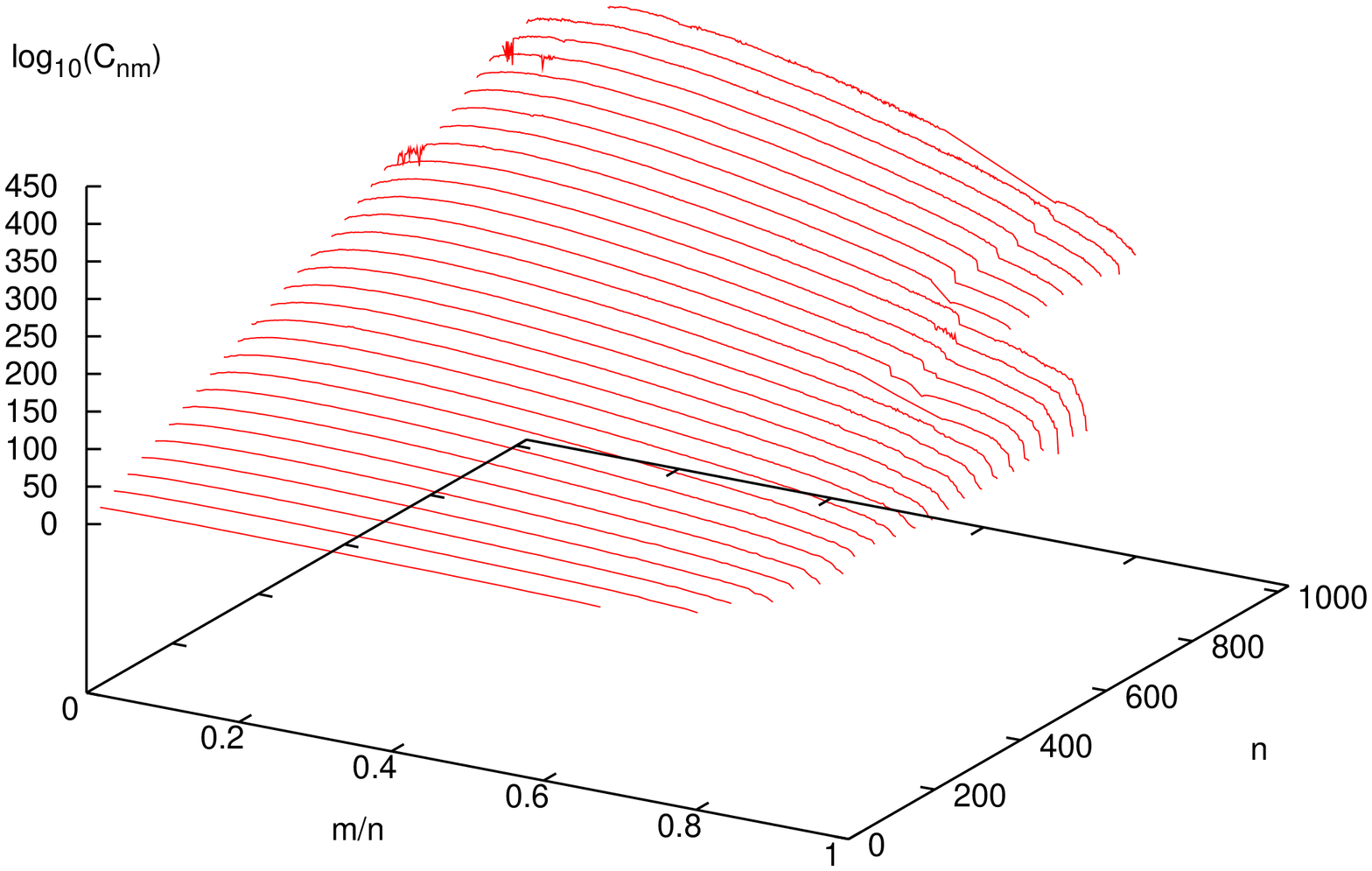}\hspace*{-1.5cm}
\includegraphics[width=5.0cm,angle=-90]{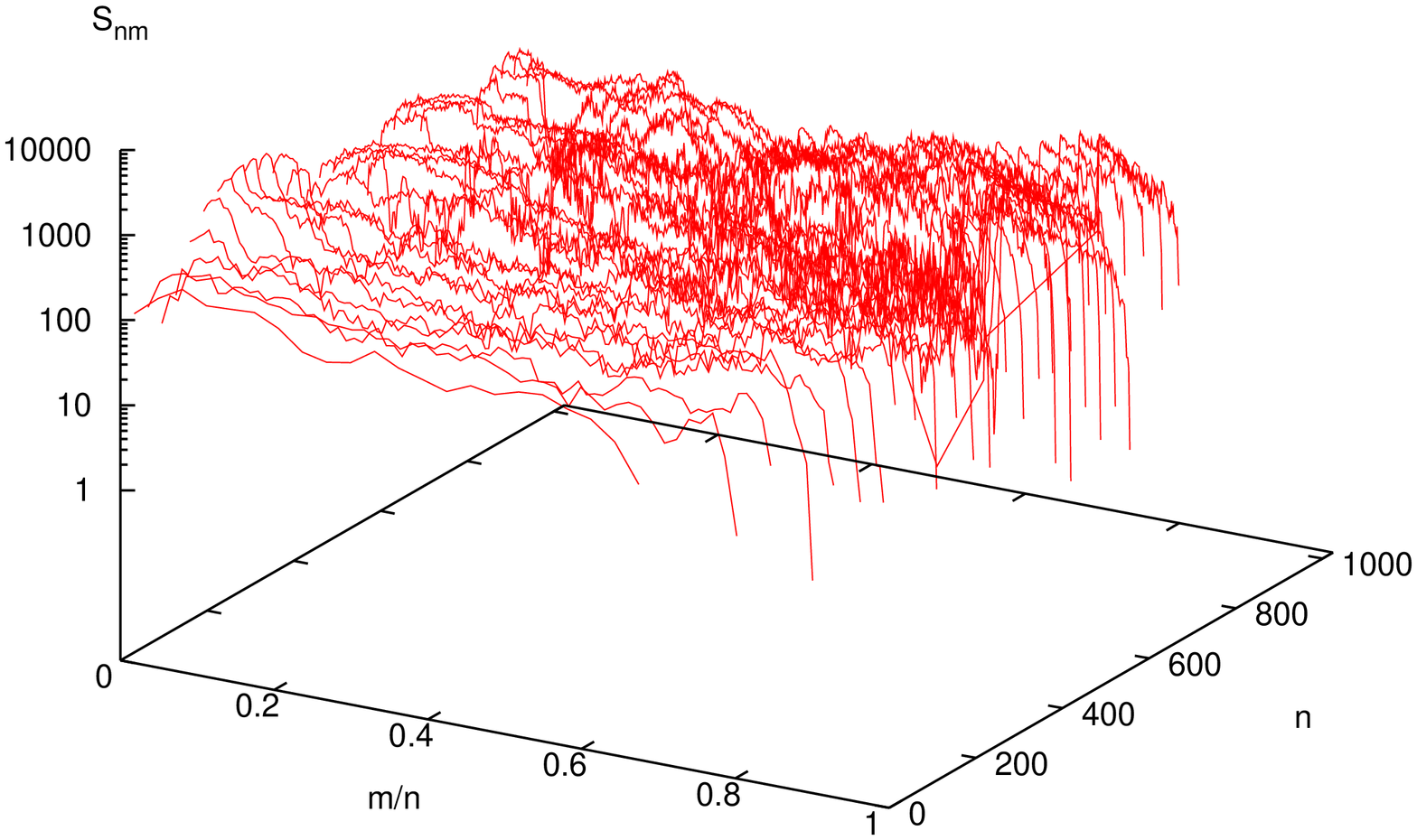}
\caption{Number of configurations $C_{n,m}$ (left) and number of generated samples (right) 
  versus internal energy $m/n$ and length $n$ for ISAW on the square
  lattice, for various total sample sizes: 
  $10^6$ (top),  $10^7$ (middle), and $10^8$ (bottom).
\label{figure10}
}
\end{figure}

We now turn to the simulation of interacting self-avoiding walks (ISAW) on
the square and simple cubic lattices.  In both cases have we simulated
walks up to length $1024$. Here, we encounter a small additional
difficulty; when PERM is initially started it is effectively blind and may
produce poor estimates of $C_{n,m}$ and this may in turn lead to overflow
problems. It is therefore necessary to stabilize the algorithm by delaying
the growth of large configurations. For this, it suffices to restrict the
size of the walks by only allowing them to grow to size $n$ once $t=cn$
tours (the number of times the algorithm returns to an object of zero size)
has been reached. We found a value of $c\approx0.1$ sufficient.

Fig.~\ref{figure10} shows the equilibration of the algorithm due to
the delay. Snapshots are taken after $10^6$, $10^7$, and $10^8$
generated samples. While the sample histogram looks relatively rough
(even on a logarithmic scale) the density of states is already rather
well behaved. In the plots one clearly sees the effect of large
correlated tours in which large number of enrichments produce many
samples with the same initial walk segment.

\begin{figure}
\begin{center}
\includegraphics[width=5.5cm]{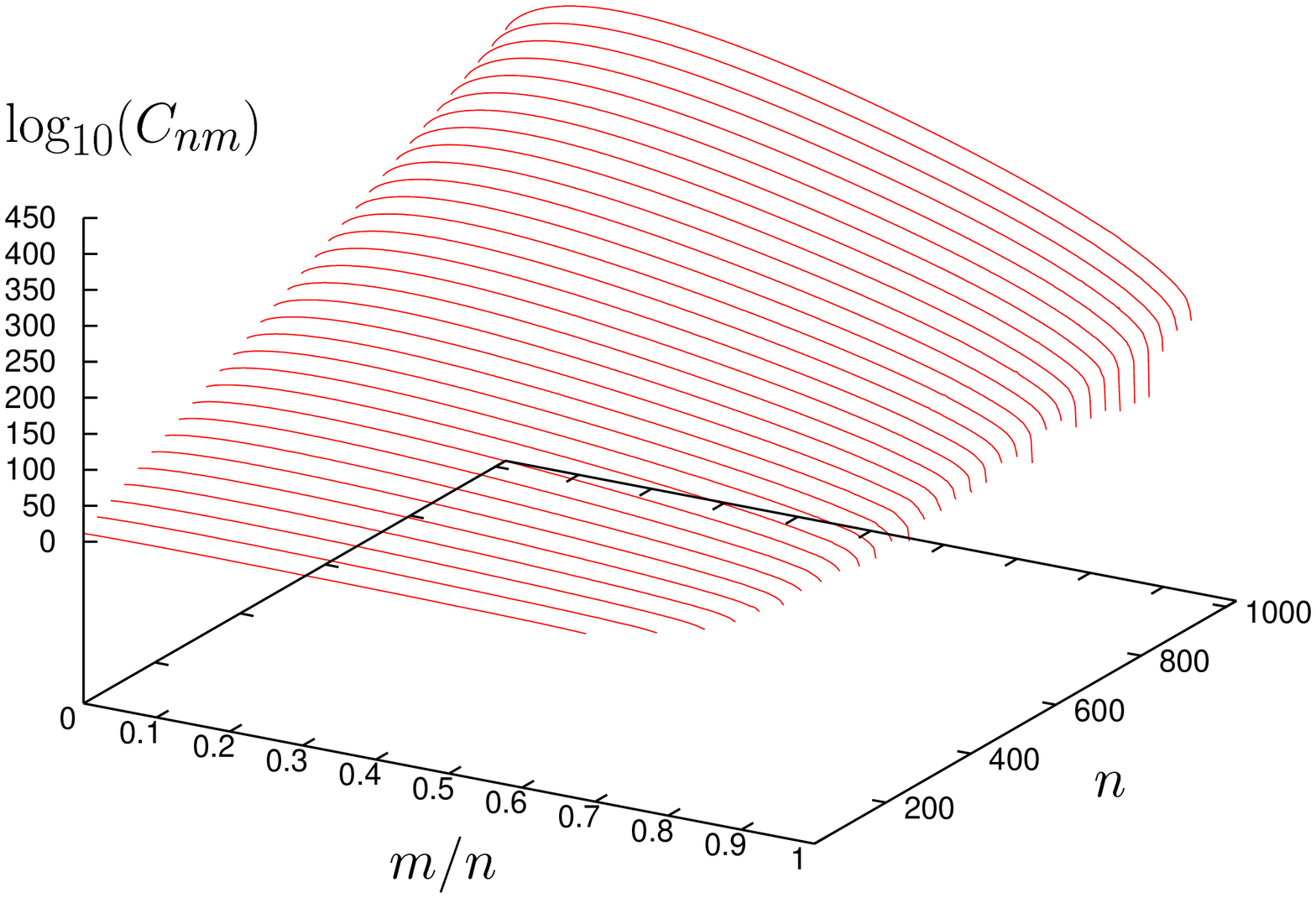}
\includegraphics[width=5.5cm]{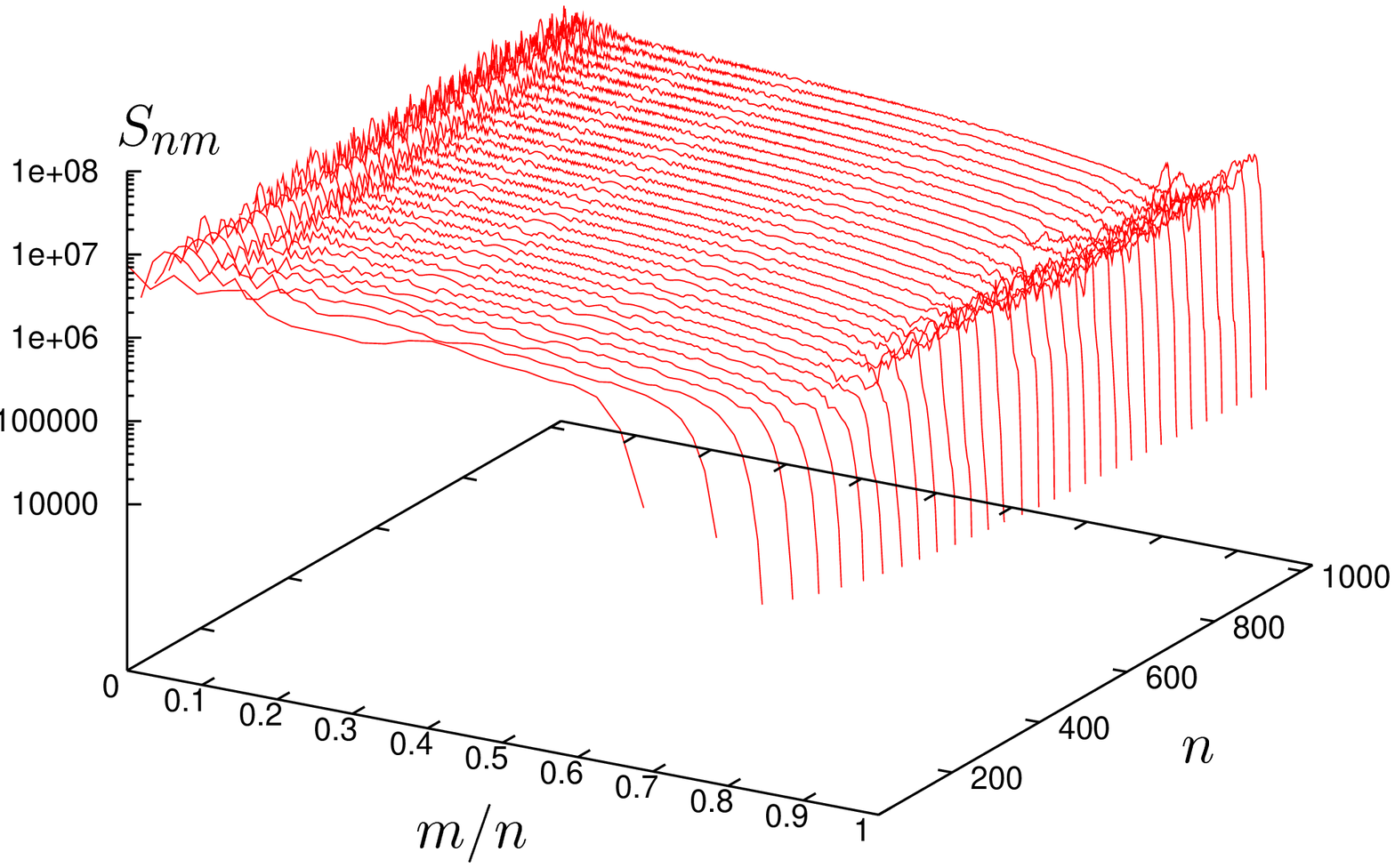}\vspace{0.5cm}\\
\includegraphics[width=5.5cm]{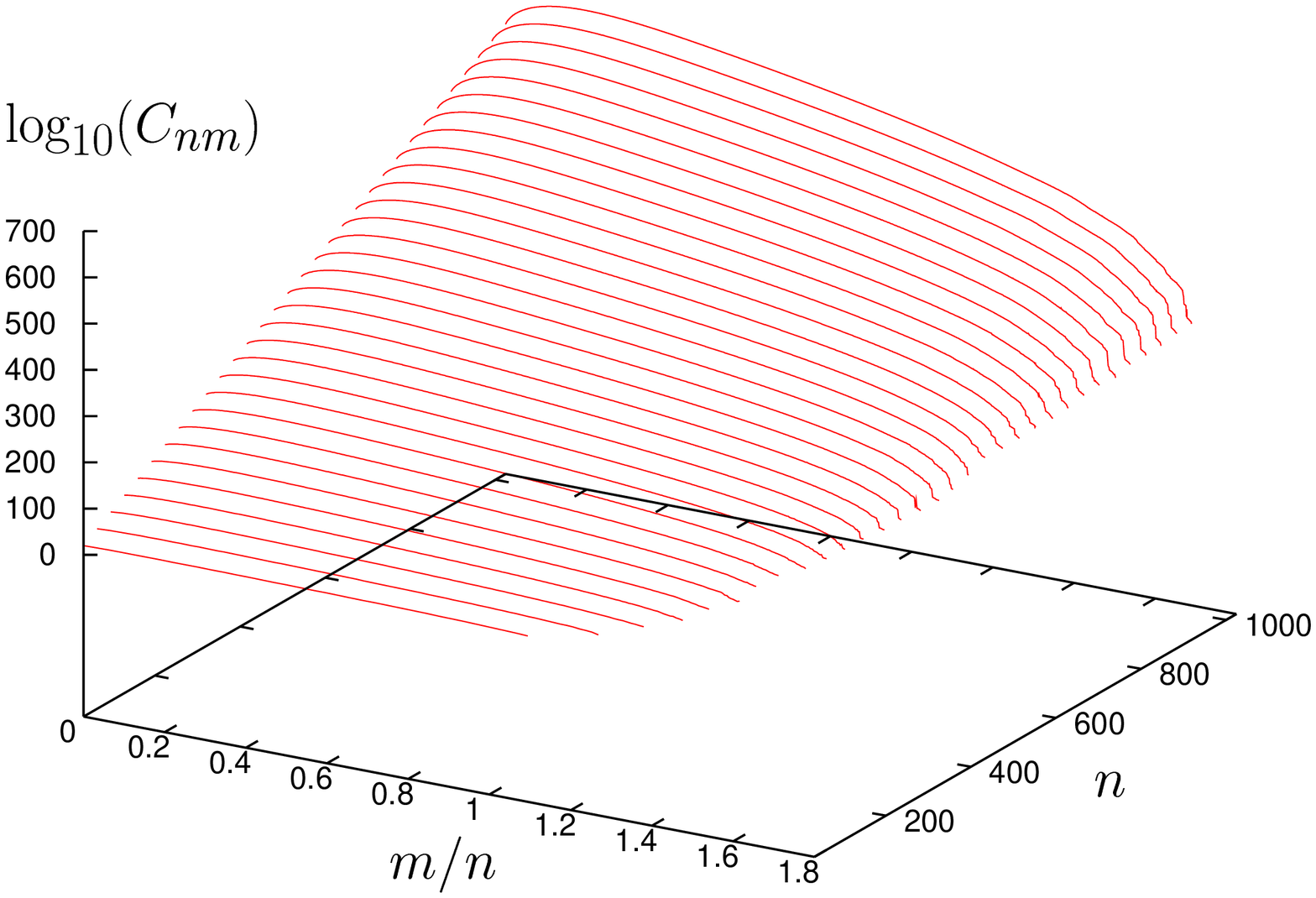}
\includegraphics[width=5.5cm]{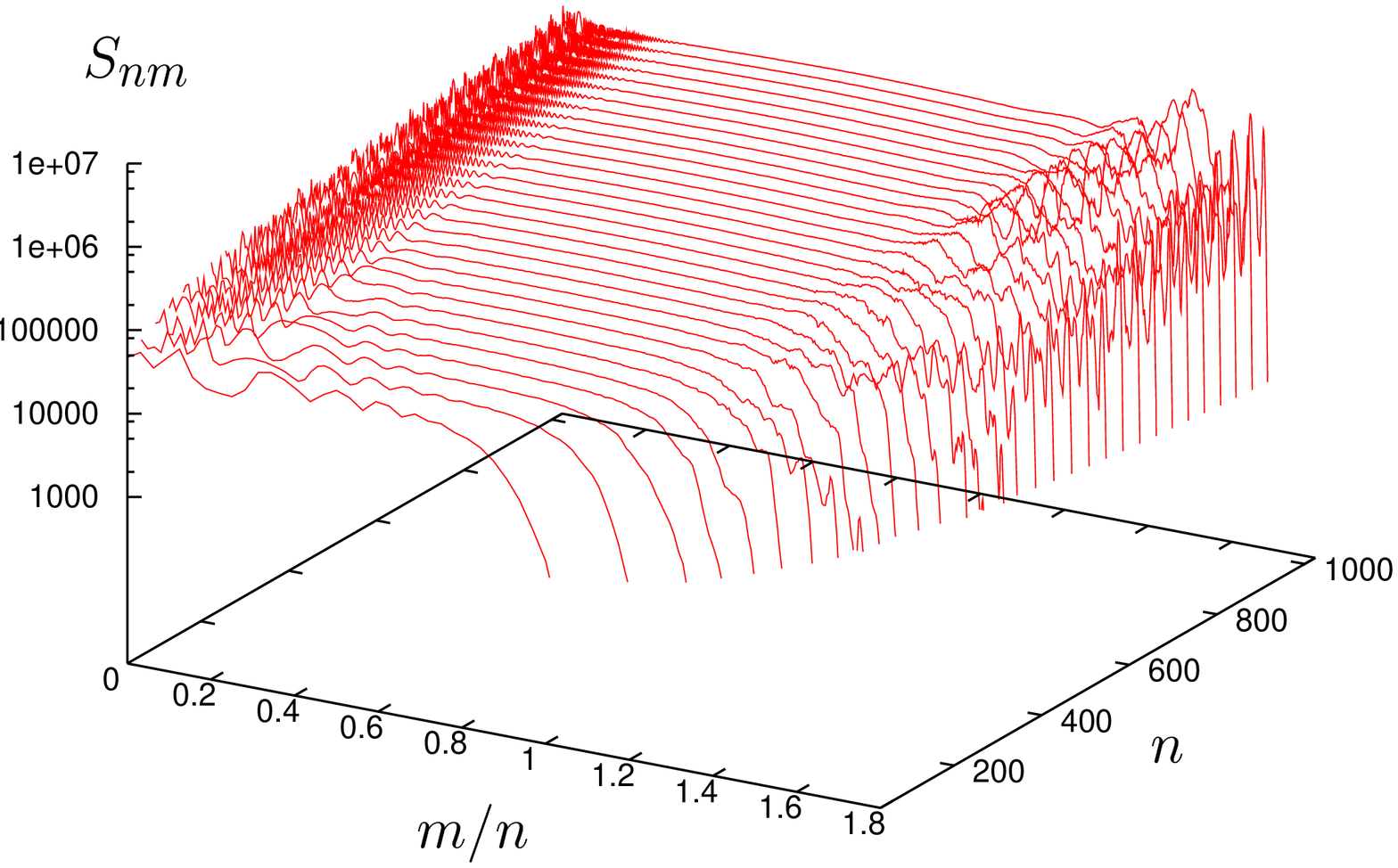}
\end{center}
\caption{Number of configurations $C_{n,m}$ (left) and number of generated samples (right)
  versus internal energy $m/n$ and length $n$ for ISAW on the square
  lattice (above) and simple cubic lattice (below)
\label{figure11}
}
\end{figure}

The final result of our simulations for interacting self-avoiding
walks in two and three dimensions is shown in Fig.~\ref{figure11}. It
clearly shows the strength of flatPERM: with one single simulation can
one obtain a density of states which ranges over more than 300 orders
of magnitude!

\begin{figure}
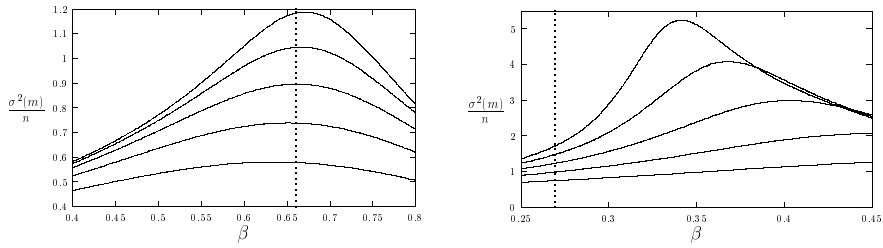

\includegraphics[width=5.5cm]{2dcv.epsi}\hspace{0.5cm}
\includegraphics[width=5.5cm]{3dcv.epsi}
\caption{
  Normalized fluctuations $\sigma^2(m)/n$ versus inverse temperature
  $\beta=1/k_BT$ for ISAW on the square lattice (above) and the simple
  cubic lattice (below) at lengths 64, 128, 256, 512, and 1024.  The curves
  for larger lengths are more highly peaked. The vertical lines denote the
  expected transition temperature at infinite length.
\label{figure12}
}
\end{figure}

From these data one can now, for example, compute the specific heat
curves $C_n=k_B(\beta\epsilon)^2\sigma^2(m)/n$. The results for both
systems are shown in Fig.~\ref{figure12}. We see that the data is well
behaved well into the collapsed low-temperature regime.

\section{Conclusion and Outlook}

We have reviewed stochastic growth algorithms for polymers. Describing 
the Rosenbluth and Rosenbluth method as an approximate counting method has enabled
us to present a straight-forward extension of simple PERM to flat histogram
PERM. Using this algorithm one can obtain the complete density of states
(even over several hundred orders of magnitude) from one single simulation.

We demonstrated the strength of the algorithm by simulating self-avoiding
walks in a strip, the HP-model of proteins, and interacting self-avoiding
walks in two and three dimensions as a model of polymer collapse.

Further applications are in preparation, \emph{eg} simulations of
branched polymers, and simulations of higher-dimensional densities of
states.

\end{document}